# Highlights

**Single-Cycle Multidirectional EOG Classification Faster than Human Reaction Time for Wearable Human-Computer Interactions**

Tasnia Nabiha, Orthy Toor, Wakim Sajjad Sakib, Abdullah Bin Shams

- Achieved highly accurate ($\approx$ 99%) classification of 10 eye-movement classes using single-cycle EOG signals.

- Developed low-latency ANN and CNN models with response times respectively, 38.6 ms and 50.82 ms, significantly below human reaction time (250 ms).

- Proposed a cascaded neural network architecture to improve multi-class classification under limited data conditions.

- Demonstrated that single-cycle EOG enables real-time, energy-efficient human–computer interaction for wearable systems.

- Introduced a latency-aware evaluation framework (CI and FoM) for balanced assessment of accuracy and responsiveness.

# Single-Cycle Multidirectional EOG Classification Faster than Human Reaction Time for Wearable Human-Computer Interactions

Tasnia Nabiha[a,1], Orthy Toor[a,1], Wakim Sajjad Sakib[b], Abdullah Bin Shams[c,*]

[a]*Department of Electrical and Electronic Engineering, Islamic University of Technology, Gazipur, 1704, Bangladesh*
[b]*Department of Biomedical Engineering, University of Cincinnati, Cincinnati, OH, 45221, USA*
[c]*Department of Electrical & Computer Engineering, University of Toronto, Toronto, Ontario, M5S 3G4, Canada*

**Abstract**

Electrooculogram (EOG) is a non-invasive bio-signal generated by the potential difference between the retina and cornea during eye movement, and is widely utilized in Human–Computer Interaction (HCI) systems. Expanding the range of detectable eye movements enhances system capability. However, increasing the number of classes typically degrades classification performance. While AI-based approaches can mitigate this limitation, their complexity increases significantly when operating on single-cycle EOG signals. Although single-cycle signals offer advantages such as low latency, reduced power consumption, and improved responsiveness, they are inherently limited by reduced informational content and higher susceptibility to noise. Ensuring low latency remains critical for real-time HCI applications, where system response must remain below human reaction thresholds. In this experimental study, using explainable AI, we address these challenges by developing 1-dimensional (1D) and cascaded ANN and CNN architectures capable of highly accurate classification across ten EOG classes (Stare, Blink, Up, Down, Right, Left, Up-left, Up-right, Down-left, and Down-right) using single-cycle signals, while simultaneously achieving latency substantially lower than human reaction time. The study achieved an accuracy of around 99% for all the models with a latency of 38.6 ms for the 1D ANN, and 82.85 ms for the cascaded CNN. These findings confirm that cascaded neural network architectures, can effectively balance high classification accuracy and low latency for single-cycle, multi-class EOG-based HCI systems under limited data availability.

## 1. Introduction

The global smart wearables market has experienced strong growth over the past decade, driven by rising health awareness, advances in miniaturized sensors, and the integration of artificial intelligence into compact, user-centric devices as shown in Fig. 1. In 2024, the global wearables market was valued at approximately USD 179.8 billion and is projected to reach nearly USD 995.2 billion by 2034, growing at a compound annual growth rate (CAGR) of 18.9% [1]. This expansion reflects increasing reliance on wearables across healthcare, fitness, consumer electronics, defense, and human–computer interaction (HCI), where continuous monitoring and real-time feedback are essential [1, 2]. Beyond wrist-worn devices, eyewear and head-wear are also expected to grow rapidly, driven by augmented and virtual reality (AR/VR), hands-free interaction, and health-sensing applications [3, 4, 5]. In particular, healthcare has emerged as a major driver, with the wearable medical devices market valued at USD 103.04 B in 2025 and projected to reach USD 505.28 B by 2034, highlighting accelerating demand for continuous monitoring and remote-care technologies [6].

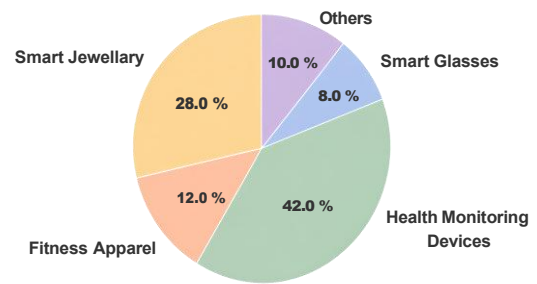

Figure 1: Global market share of different smart wearable devices

Modern wearable systems increasingly rely on advanced human–computer interaction (HCI), particularly in medical assistive devices for individuals with motor impairments and in emerging augmented and virtual reality (AR/VR) platforms [7]. These applications demand highly responsive interfaces, as interaction speed directly influences usability and user satisfaction [8]. To support such interaction, recent systems increasingly utilize biosignals such as electroencephalography (EEG), electromyography (EMG), and electrooculography (EOG) [9, 10, 11]. Among these modalities, EOG has emerged as a particularly promising technology due to its non-invasive and energy-efficient ability to capture eye movements [12]. EOG measures variations in the corneo-retinal potential as shown in Fig. 2, typically ranging from tens of microvolts to several millivolts, which can be reliably detected using surface electrodes placed around the eyes. [13]

---

*Corresponding author.
 Email addresses:* tasnianabiha@iut-dhaka.edu (Tasnia Nabiha), orthytoor@iut-dhaka.edu (Orthy Toor), sakibwd@mail.uc.edu (Wakim Sajjad Sakib), ab.shams@utoronto.ca (Abdullah Bin Shams)
 [1]These authors contributed equally to this work.

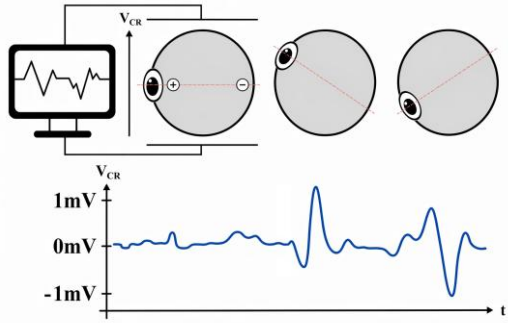

Figure 2: Corneo-retinal potential during eye movements

This capability has enabled EOG-based wearables to find applications across a wide range of domains, including augmented and virtual reality (AR/VR) systems for hands-free interaction, autonomous and semi-autonomous vehicles for driver attention and fatigue monitoring, remote and contactless device control, and assistive technologies for individuals with severe motor impairments or paralysis [14, 15]. Beyond these, EOG wearables are increasingly explored in smart conferencing environments, gaming and immersive entertainment, rehabilitation systems, sleep assessment, and smart lifestyle applications, where intuitive, eye-driven interaction can significantly enhance user experience [16]. The growing adoption of EOG in wearable ecosystems reflects its unique balance of robustness, low computational overhead, and suitability for real-time human–computer interaction.

As eye-based interaction becomes increasingly integral to wearable systems, the need for precise and reliable control under strict latency constraints becomes a central challenge. In wearable assistive applications, accurate and timely classification of EOG signals is essential to enable real-time human-computer interaction (HCI). For real-time operation, EOG signals must be processed within the average human reaction time of 200–300 ms [17] to ensure seamless interaction. In applications such as assistive robotics, autonomous and semi-autonomous systems, AR/VR interaction, and smart mobility, decision-making within or below this temporal window is essential for effective and safe operation. However, most existing EOG classification frameworks rely on multi-cycle signal analysis, in which multiple consecutive EOG cycles are aggregated to capture temporal dependencies before producing a final decision [18, 19, 20, 21]. While this approach improves robustness and classification stability, it inherently introduces additional processing delay, resulting in latency that is often incompatible with real-time wearable applications. Consequently, shifting toward single-cycle EOG analysis, in which classification is performed within a single ocular cycle, provides a direct mechanism for latency reduction, lower power consumption, and real-time system responsiveness. Moreover, current EOG-based systems largely focus on a limited set of eye movements, typically horizontal, vertical, and blink detection, which restricts interaction expressiveness and reduces flexibility in more complex wearable and assistive scenarios.

Although single-cycle EOG analysis offers a direct mechanism for reducing latency, power consumption, and response time, it introduces significant challenges. Unlike multi-cycle approaches that benefit from temporal averaging, a single ocular cycle contains limited information, making classification more sensitive to noise as demonstrated in Fig. 3, baseline drift, motion artifacts, and inter-user variability. As the number of eye-movement classes increases, discriminative capability further decreases, leading to higher ambiguity and false activations. Feature extraction from such short temporal windows is inherently less reliable, particularly in real-world wearable settings where bioelectrical signals are non-stationary. Nevertheless, the growing demand for low-latency and highly responsive systems has motivated the adoption of data-driven and AI-based approaches capable of extracting discriminative patterns from limited and noisy signals. [22, 23]

Motivated by these challenges, this work focuses on enabling accurate and low-latency eye-movement recognition from two-channel, single-cycle EOG signals for practical wearable applications. EOG data were collected for a diverse set of eye movements, including cardinal and diagonal directions as well as blink activity, using a minimal and wearable-friendly electrode configuration. The acquired signals were processed to mitigate noise and artifacts while preserving discriminative information within a single ocular cycle. Artificial intelligence–based models were then developed to perform reliable classification without relying on multi-cycle aggregation. From the survey of the literature in Section 2 and Table 1, most studies employ classical Machine Learning (ML) algorithms, typically limited to eight or fewer eye movement classes and relying on multi-cycle EOG signals, achieving accuracies around 95%. While effective for simpler settings, these methods struggle with increased class diversity and reduced temporal information. In contrast, our work considers a larger number of classes using a single-cycle approach, which increases classification complexity. To address this, we employ neural networks, which can automatically learn hierarchical features from raw inputs and capture subtle variations in EOG signals. Both Artificial Neural Networks (ANN) and Convolutional Neural Networks (CNN) are implemented to compare their performance. Furthermore, we investigate two architectural strategies based on data availability: a 1D CNN for sufficiently large datasets and a cascaded architecture for limited data scenarios. The cascaded approach decomposes complex multi-class classification into simpler stages, improving accuracy while maintaining low computational cost and latency suitable for real-time wearable systems, as shown in Table 1.

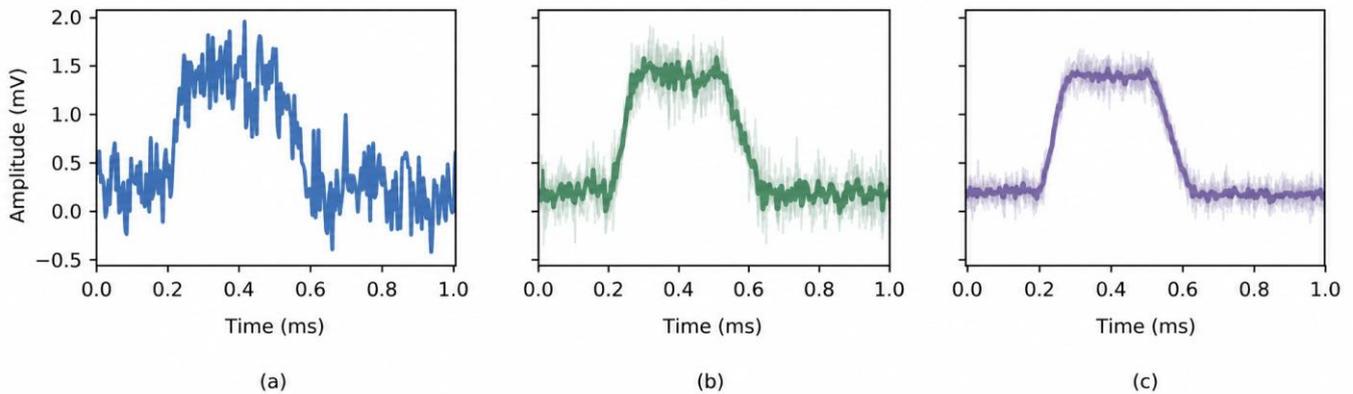

Figure 3: Comparison of EOG signal quality with cycle averaging. Noise reduction and SNR improvement are observed as the number of averaged cycles increases: (a) single cycle (b) 5-cycle average, and (c) 10-cycle average showing clearer and more stable signal characteristics.

## 2. Literature Review

Extensive research has investigated electrooculography (EOG)-based systems for enhancing wearable human–computer interaction, with most approaches relying on multi-cycle signal acquisition and a limited set of eye-movement classes. Early studies primarily focused on simple directional control using handcrafted features and conventional machine learning models. For instance, Rusydi et al. [24] employed two-channel EOG signals acquired through a custom analog front-end consisting of amplification and filtering stages, followed by digitization using an NI USB-6008 data acquisition board. The authors extracted the area under the EOG signal from both channels, forming a two-dimensional feature space, and used a multiclass Support Vector Machine (SVM) to classify four cardinal eye movements (left, right, up, and down). The system achieved an overall accuracy of 99%, with perfect precision for horizontal movements and slightly lower performance (96%) for vertical movements. However, reduced robustness in the horizontal channel was reported due to electrode placement asymmetry at the outer canthi, emphasizing a fundamental limitation in EOG acquisition that affects signal consistency.

To enhance feature representation, subsequent studies incorporated more advanced signal processing techniques. Hernández et al. [20] utilized wavelet transform-based feature extraction combined with a comprehensive set of statistical features, including RMS, amplitude, variance, covariance, median, average, and power spectrum. Using supervised classifiers such as KNN, SVM, and Decision Tree, they classified five eye movements (left, right, up, down, and blink), achieving accuracies of 69.4%, 76.9%, and 60%, respectively. Despite employing richer features, the results revealed a consistent trend: vertical eye movements were classified more reliably than horizontal ones. This behavior is attributed to stronger and more stable signal amplitudes during vertical gaze shifts, whereas horizontal movements exhibit mirror symmetry and are more susceptible to electrode misplacement and muscle artifacts, leading to overlapping feature distributions and reduced separability.

To increase interaction flexibility, later works expanded the number of detectable eye gestures. Lin et al. [25] classified nine eye movement classes, including diagonal directions, using a four-channel EOG system. Signal acquisition was performed using an EOG Mindo device with high Common Mode Rejection Ratio (CMRR) achieved via an INA2126 instrumentation amplifier. The preprocessing pipeline included a 0.1 Hz high-pass filter to remove baseline drift, a moving average filter for noise reduction, and a synchronization filter to suppress frequencies above 62.5 Hz. Blink artifacts were detected and removed using slope-based methods, and peak values were used for classification. While the system achieved an overall accuracy of 94.67%, performance degradation was observed for diagonal movements, which can be explained by the superposition of horizontal and vertical signal components, reducing class separability compared to cardinal directions.

In parallel, several studies explored low-complexity and embedded implementations. Pratomo et al. [26] developed a rule-based classification system using two-channel EOG signals acquired via AD620 amplification and filtered within a 0.1–30 Hz band. The signals were digitized using the Arduino Uno's ADC modules, and classification was performed using threshold-based logic without explicit feature extraction. While horizontal movements achieved 100% accuracy, vertical classification showed significant imbalance, with downward movements achieving 96% accuracy and upward movements only 64%. This disparity indicates that peak-based features are insufficient for robust vertical classification, likely due to residual blink artifacts and baseline drift. Additionally, the center (no-movement) class achieved only 84% accuracy, highlighting the difficulty of distinguishing intentional inactivity from small involuntary eye movements.

Similarly, Kabir et al. [27] developed a low-cost EOG-based cursor control system using Ag–Ag/AgCl electrodes and a custom acquisition circuit with AD620 amplification and second-order filtering (0.67 Hz high-pass and 33.33 Hz low-pass). Feature extraction was limited to rolling mean, Fast Fourier Transform (FFT), and gradient-based descriptors. Using SVM and

Table 1: Comparison of Existing EOG-Based Eye Movement Classification Studies

| Ref | Model | Classes | Cycle | Channel | Accuracy | Latency | CI | FoM |
|---|---|---|---|---|---|---|---|---|
| [20] | KNN, SVM, DT | 5 [Left, Right, Up, Down, Blink] | Multi | 2 | 69.4% (KNN), 76.9% (SVM), 60% (DT) | N/A | 3.56 | – |
| [24] | SVM | 4 [Right, Left, Up, Down] | Multi | 2 | 99% | N/A | 3.95 | – |
| [25] | ML-based classifier | 8 [Up, Down, Left, Right, Upleft, Upright, Downleft, Downright] | Multi | 4 | 94.67% | N/A | 7.51 | – |
| [26] | Threshold + ML | 5 [Right, Left, Up, Down, Forward] | Multi | 2 | Error-based performance | N/A | – | – |
| [27] | SVM, MLP | 5 [Up, Down, Left, Right, Blink (intent/unintended)] | Multi | 2 | 93% | N/A | 4.56 | – |
| [28] | GRU, Bi-GRU | 4 [Up, Down, Left, Right] | Multi | 2 | 99.77% (V), 99.74% (H) | N/A | 3.99 | – |
| [29] | ResNet-50 | 6 [Right, Left, Up, Down, Center, Double Blink] | Multi | 2 | 95.8% | N/A | 5.70 | – |
| [30] | Direction Detection Algorithm | 4 [Right, Left, Upward, Downward] | – | 2 | 83% | 218 ms | 3.09 | 3.09 |
| This study | ANN | 10 [Blink, Stare, Up, Down, Left, Right, Upleft, Upright, Downleft, Downright] | Single | 2 | 98.85% | 38.6 ms | 9.87 | 9.87 |
| | Cascaded ANN | | | | 99.30% | 99.72 ms | 9.92 | 9.92 |
| | 1D CNN | | | | 97.35% | 50.82 ms | 9.71 | 9.71 |
| | Cascaded CNN | | | | 99.75% | 82.25 ms | 9.97 | 9.97 |

*Note:* CI = Classification Index; FoM = Figure of Merit.

Multilayer Perceptron (MLP) classifiers, the system achieved accuracies of 93% and 80%, respectively, across six classes, including intentional and unintentional blinks. However, performance variability across subjects was significant due to electrode placement inconsistencies, physiological differences, and limited feature representation, all of which contributed to reduced generalization capability.

More recent work has shifted toward deep learning approaches to automate feature extraction and improve classification performance. Roy et al. [28] applied recurrent neural networks, including GRU and bidirectional GRU, using features such as peak amplitudes, RMS, variance, and zero-crossing rate. After applying a fourth-order Butterworth low-pass filter, the models achieved accuracies exceeding 99.7% for four directional eye movements. Similarly, Reda et al. [29] employed convolutional architectures, including ResNet-50, on filtered and augmented EOG signals to classify six eye movements, achieving up to 95.8% accuracy. These models leverage deep hierarchical feature extraction to improve robustness; however, their performance is strongly influenced by the restriction to simpler class sets and multi-cycle temporal windows, which inherently simplify the classification task.

Despite these advancements, several limitations remain evident across the literature. First, most studies focus predominantly on cardinal eye movements, which produce distinct and easily separable signal patterns, resulting in consistently high but less challenging performance metrics. In contrast, diagonal eye movements are less explored, despite being essential for richer and more natural interaction, due to their overlapping signal characteristics and reduced separability. Second, although classification accuracy is widely reported, system latency is rarely quantified, even though it is critical for real-time wearable applications. For example, Joko et al. [30] reported a latency of 218 ms, which approaches the average human reaction time (∼250 ms), potentially limiting responsiveness in interactive systems.

Overall, prior work demonstrates that high classification accuracy is achievable under controlled conditions; however, these systems often rely on multi-cycle inputs, limited class diversity, or overlook real-time constraints. These limitations highlight the need for EOG-based systems capable of supporting a broader set of eye movements, including diagonals, while operating under low-latency, single-cycle conditions suitable for real-time interaction.

Beyond these design limitations, evaluation practices also remain inconsistent across studies. Although classification accuracy is widely reported for varying numbers of eye-movement classes, direct comparison is difficult due to differences in class complexity and experimental setups [31]. Furthermore, system latency is rarely quantified despite its importance for real-time usability, making accuracy alone an incomplete measure of performance.

To address these limitations, we introduce two unified performance metrics. First, the Classification Index (CI) normalizes accuracy with respect to both random chance and the number of classes:

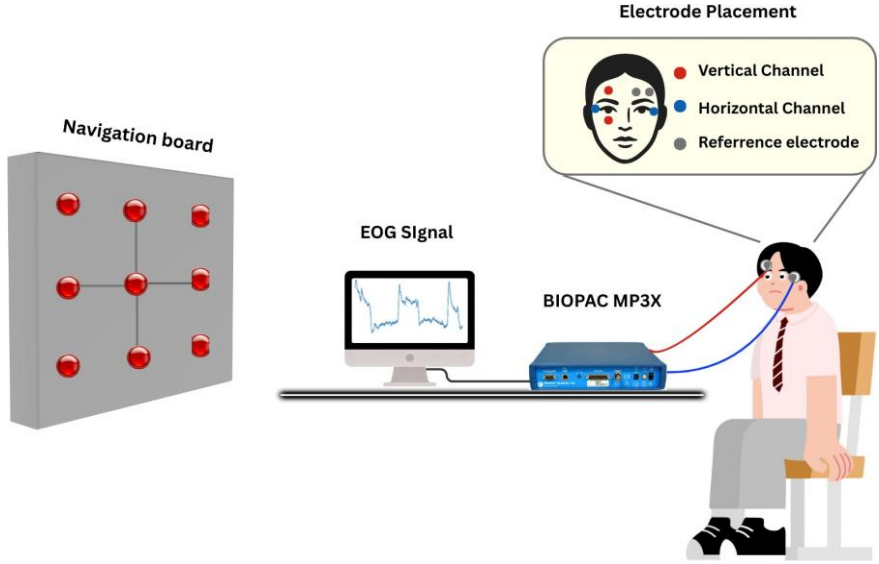

Figure 4: Experimental data acquisition setup showing electrode placement, BIOPAC MP3X recording system, and LED-based visual cue synchronization mechanism used for controlled EOG data collection.

$$\text{CI} = k \cdot \frac{\text{Classification Accuracy} - \frac{1}{k}}{1 - \frac{1}{k}} \quad (1)$$

where $k$ is the number of classes, and classification accuracy is between 0-1. Here, $1/k$ represents the baseline accuracy of random guessing, and normalization by $1 - 1/k$ scales performance relative to the maximum achievable improvement. The multiplication by $k$ compensates for the bias that systems with fewer classes tend to achieve higher accuracy. Thus, CI provides a class-normalized measure of classification performance, enabling fair comparison across studies with different class counts.

Second, to incorporate real-time constraints, we define a latency-aware Figure of Merit (FoM) as:

$$\text{FoM} = \text{CI} \cdot min(1, \frac{250 \text{ ms}}{t_{sys}}) \quad (2)$$

where 250 ms is the average human reaction time and $t_{sys}$ is the system latency in ms. This unitless metric evaluates performance relative to the average human reaction time. To avoid over-rewarding extremely low latency, the ratio is capped at 1 when system latency is below human reaction time. Consequently, FoM prioritizes systems that are both accurate and responsive, while penalizing delays that exceed human perceptual limits.

Since latency is rarely quantified in existing work, such composite evaluation remains largely absent in the literature. However, these metrics enable a consistent and physiologically meaningful comparison framework for EOG-based wearable systems, and can be directly applied to the proposed approach.

## 3. Methodology

### 3.1. Data Collection

Electrooculography (EOG) signals corresponding to ten voluntary eye movement classes (stare, blink, left, right, up, down, up-left, up-right, down-left, and down-right) were acquired from a group of 20 healthy participants. The participant pool was balanced in terms of gender to minimize bias in physiological variability. All recordings were conducted in a controlled environment at the KUET Biomedical Laboratory to ensure signal consistency and minimal external interference.

Data acquisition was performed as shown in Fig. 4 using the BIOPAC MP3X data acquisition system. Disposable electrodes were placed around the eyes using electrolyte gel to reduce electrode–skin impedance. A total of six electrodes were used, where four electrodes captured horizontal and vertical eye movements, and two electrodes served as reference points.

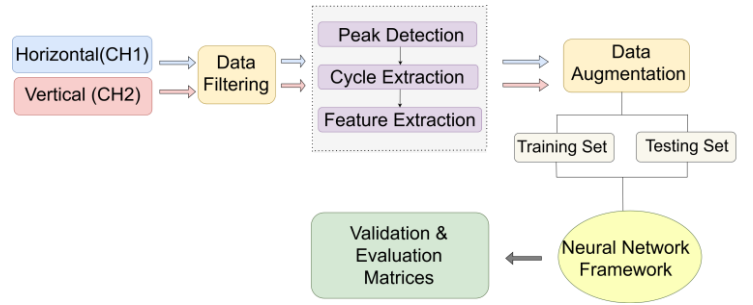

Figure 5: Workflow of the Study

The EOG signals were sampled at a frequency of 125 Hz and monitored in real time on a desktop computer connected to the acquisition unit. Visual cues were provided using an LED-based stimulus system controlled by an Arduino board to guide participants in performing the required eye movements. Each participant completed multiple trials for every eye movement class, with each trial lasting 20 seconds as described in [30].

In total, 450 EOG signal samples were collected across all participants. The recorded data were stored digitally for subsequent preprocessing, feature extraction, and classification.

*3.2. Signal Preprocessing*

To effectively capture the temporal characteristics of eye movements, the recorded EOG signals were segmented using a fixed window length of 350 samples, which was empirically selected to represent a complete eye movement cycle. During preliminary inspection, the raw signals exhibited baseline drift, motion artifacts, and high-frequency noise that significantly masked the underlying eye movement patterns. To mitigate these issues, a two-stage signal preprocessing framework was developed through extensive experimentation.

*3.2.1. Butterworth High-Pass Filter*

In the first stage, baseline drift and slow-varying offsets were removed using a Butterworth high-pass filter. This step, commonly referred to as detrending, suppresses low-frequency components while preserving the dynamic features associated with eye movements. The Butterworth filter was chosen due to its smooth frequency response and minimal phase distortion within the passband. Frequencies below the selected cutoff were significantly attenuated, resulting in a stabilized signal baseline suitable for further analysis.

$$H_{\text{high-pass}}(j\omega) = \frac{j\omega}{j\omega + \omega_c} \quad (3)$$

where $\omega$ is the angular frequency and $\omega_c$ is the cutoff angular frequency.

*3.2.2. Wavelet Denoising*

Following drift removal, wavelet-based denoising was applied to reduce high-frequency noise and transient artifacts. This technique decomposes the signal into multiple resolution levels using wavelet transform, allowing noise components to be isolated in specific frequency bands. Noise-dominated wavelet coefficients were selectively thresholded, and the signal was subsequently reconstructed using the remaining coefficients. This approach effectively suppressed unwanted noise while retaining critical signal features such as peaks and transitions corresponding to different eye movements.

$$\hat{x}(t) = \sum_j w_j \cdot \varphi_{j,k}(t) \quad (4)$$

where $\hat{x}(t)$ denotes the denoised signal, $w_j$ represent the wavelet coefficients, and $\varphi_{j,k}(t)$ are the corresponding wavelet basis functions.

*3.3. Peak Detection and Cycle Extraction*

After preprocessing, a peak detection algorithm was applied to identify individual eye-movement cycles within the EOG signals. A custom function was designed to detect peaks by tracking signal amplitudes that exceeded a predefined threshold in both upward and downward directions, thereby determining the onset and offset of each ocular movement. The midpoint between these two threshold-crossing points was defined as the peak location for the corresponding cycle.

To improve detection reliability and reduce false positives caused by noise or involuntary eye actions, parameters such as minimum peak distance, peak height, and prominence were carefully tuned. The minimum peak distance constraint ensured that closely spaced fluctuations were not misclassified as multiple peaks, enabling more accurate cycle identification.

Each detected peak was subsequently used as a temporal reference, around which a fixed-length signal window was extracted by collecting data samples from both sides of the peak. This procedure allowed the complete temporal characteristics of a single eye-movement cycle to be isolated and standardized for subsequent analysis, feature extraction, and classification.

*3.4. Feature Extraction*

A total of 13 features, comprising statistical and gradient-based measures, were extracted from each individual EOG cycle for every recording channel. The statistical features were designed to characterize the overall amplitude behavior, variability, distribution shape, and frequency-domain properties of the signal, while the gradient-based features captured the dynamic temporal characteristics of eye movements.

*3.4.1. Statistical Features*

Seven statistical features were computed from each EOG cycle to describe its amplitude distribution and spectral content. These include:

- **Mean absolute amplitude:** Represents the average absolute value of the signal over a cycle, reflecting overall signal energy and helping to differentiate movements of varying intensity.

- **Maximum amplitude:** Indicates the peak value within a cycle, capturing the maximum ocular muscle activation and supporting class discrimination.

- **Minimum amplitude:** Denotes the lowest signal value within a cycle, complementing the maximum amplitude in defining signal range and polarity.

- **Standard deviation:** Measures the dispersion of the signal around its mean, reflecting the variability and stability of the eye movement pattern.

- **Skewness:** Quantifies the asymmetry of the signal distribution, highlighting directional bias in waveform shape useful for distinguishing movement types.

- **Kurtosis:** Describes the sharpness or peakedness of the distribution, indicating abrupt variations or outlier-like behavior in the signal.

- **Mean FFT magnitude:** Represents the average spectral energy of a cycle, capturing frequency-domain characteristics of eye movements. It is computed as the mean of the magnitude spectrum obtained from the Fast Fourier Transform (FFT) of the signal:

$$\text{Mean}_{FFT} = \frac{1}{N} \sum_{i=1}^{N} X_i \quad (5)$$

where $X_i$ denotes the magnitude of the $i^{th}$ frequency component, and $N$ is the total number of frequency bins.

### 3.4.2. Gradient-Based Features

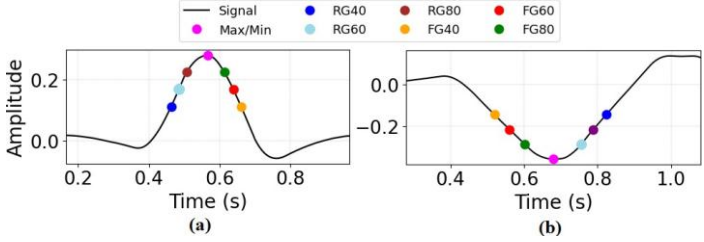

Figure 6: Identification of gradient-based features (RG and FG) at 40%, 60%, & 80% amplitude levels for both (a) positive and (b) negative signal cycles.

To capture the temporal dynamics and rate of signal variation that amplitude-based features alone cannot adequately represent, six gradient-based features were extracted. These features quantify how rapidly the signal rises and falls within a cycle and are particularly useful for distinguishing subtle differences among eye movement classes across electrode configurations.

Each EOG cycle was divided into two phases: a rising phase (from cycle onset to peak) and a falling phase (from peak to cycle offset) for positive peaks, with the phase order reversed for negative peaks. Threshold-based slope measurements were computed at three normalized amplitude levels—40%, 60%, and 80% of the peak amplitude.

For a peak amplitude $A_{peak}$ occurring at time $t_{peak}$, the gradients were defined as:

$$RG_p = \frac{p \cdot A_{peak}}{t_{peak} - t_{Rp}} \quad (6)$$

$$FG_p = \frac{p \cdot A_{peak}}{t_{Fp} - t_{peak}} \quad (7)$$

where $p \in \{40, 60, 80\}$; $t_{Rp}$ and $t_{Fp}$ denote the times when the signal crosses the threshold amplitude on the rising and falling phases, respectively.

This resulted in 6 gradient features:

- Rising Gradient at 40%, 60%, and 80% of the peak amplitude (RG40, RG60, RG80),
- Falling Gradient at 40%, 60%, and 80% of the peak amplitude (FG40, FG60, FG80).

As illustrated in Fig. 6, for a peak occurring at time $t_{peak}$, the gradients were calculated using the time instants $t_1$ and $t_2$, which correspond to the moments when the signal crosses the specified threshold amplitude during the rising and falling phases, respectively. These gradient measures effectively capture the rate of change of the EOG signal at critical points within each cycle.

Table 2: Data Segmentation

| Segment | Classes | Samples |
|---|---|---|
| Cardinal | Blink | 200 |
| | Stare | 200 |
| | Up | 200 |
| | Down | 200 |
| | Lateral | 1200 |
| Right | Right | 200 |
| | DownRight | 200 |
| | UpRight | 200 |
| | Left segment | 600 |
| Left | Left | 200 |
| | DownLeft | 200 |
| | UpLeft | 200 |

### 3.5. SMOTE Algorithm

During experimentation with the single-cycle processed EOG data, two major challenges were observed: class imbalance and limited sample availability for certain eye-movement classes. These issues can negatively affect model generalization and bias the learning process toward majority classes. To address this problem, the Synthetic Minority Oversampling Technique (SMOTE) [32] was applied.

SMOTE generates synthetic samples for minority classes by interpolating between existing data points in the feature space. This approach increases class representation without simply duplicating samples, thereby improving dataset diversity. The application of SMOTE resulted in a more balanced dataset, enabling more robust and unbiased model training across all eye movement classes [33, 22].

### 3.6. Dataset Segmentation and Splitting

To facilitate hierarchical and robust model training, the dataset was segmented into three major groups based on eye-movement characteristics: Cardinal, Left, and Right movements. The Cardinal segment included blink, up, down, and combined lateral movements, while the Left and Right segments consisted of their respective directional eye movements. The distribution of samples across these segments is summarized in Table 2.

After segmentation, each segment was divided into training and validation sets. A stratified split was employed, where 80% of the data were used for model training and the remaining 20% were reserved for validation. This partitioning strategy ensured reliable performance evaluation while maintaining sufficient training data for learning discriminative patterns.

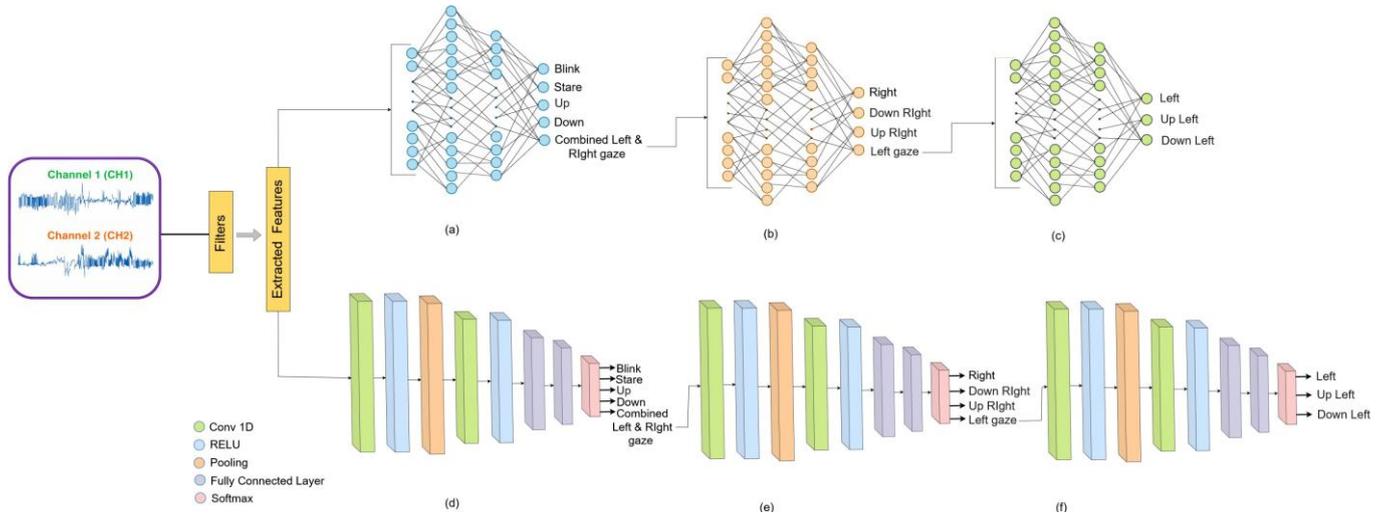

Figure 7: Cascaded module architectures (a) ANN-1 (b) ANN-2 (c) ANN-3 (d) CNN-1 (e) CNN-2 (f) CNN-3.

### 3.7. Artificial Neural Network (ANN) and Convolutional Neural Network (CNN)

An Artificial Neural Network (ANN) is a machine learning model inspired by the neural structure of the human brain [34, 35, 36]. It consists of interconnected neurons organized into input, hidden, and output layers. The input layer receives raw data, the hidden layers learn and extract complex patterns, and the output layer produces the final prediction.

In contrast, a Convolutional Neural Network (CNN) is a deep learning architecture motivated by the human visual processing system and is specifically designed for pattern recognition tasks. CNNs employ a specialized structure that leverages weight sharing and hierarchical feature extraction [37, 38], significantly reducing the number of trainable parameters and enabling faster computation, improved convergence, and efficient decision-making [13]. A typical CNN architecture includes several functional layers: the input layer accepts data, convolutional layers apply filters to extract meaningful features, and pooling layers reduce the spatial dimensions of feature maps. These features are then processed by fully connected layers for classification, and the output layer generates the final result. Additionally, CNNs often incorporate normalization layers to enhance training stability and convergence, as well as activation layers that introduce nonlinearity into the network, such as ReLU, sigmoid, and tanh functions.

Classical machine learning classifiers such as Support Vector Machine (SVM), K-Nearest Neighbor (KNN), Decision Tree (DT), and Linear Discriminant Analysis (LDA) have been widely used in EOG classification tasks. However, these methods generally depend on handcrafted feature extraction and careful feature selection [20], which may fail to capture subtle temporal variations and nonlinear dependencies present in biosignals. Their performance often degrades when dealing with noisy data, inter-subject variability, and high-class multi-class problems, especially when only short-duration signal segments are available [39]. Since single-cycle signals contain limited information and are more susceptible to noise, automated feature learning becomes particularly important. For this reason, ANN and CNN based approaches were adopted in this study, as they can learn discriminative representations directly from the signal while providing stronger generalization capability in multiclass scenarios [40].

Furthermore, a cascaded framework was explored to address the increased complexity associated with ten-class classification. Instead of forcing a single model to separate all classes simultaneously, the cascaded strategy decomposes the task into multiple simpler stages, thereby improving classification accuracy, reducing confusion among similar gaze patterns, and maintaining low computational latency suitable for real-time HCI applications.

### 3.8. Cascaded Architecture

To develop the cascaded modules for both ANN and CNN, a unified architectural design was adopted for both models, as illustrated in Fig. 7. The first-stage model performs classification of the four primary classes derived from composite gaze patterns. If this model predicts the class as Combined Left & Right Gaze, a second-stage model is subsequently activated to further classify the four sub-classes corresponding to right-gaze movements. Alternatively, if the prediction corresponds to Left Gaze, a third-stage model is invoked to discriminate among the three classes specific to left-gaze movements.

### 3.9. Cross-Validation and Evaluation Metrics

Cross-validation is the technique where the dataset is divided into several parts and the model is run on each part, then repeated again for other parts and the results are averaged. These parts are called folds. In this study we used 5-fold cross validation. To evaluate the model's performance, we used four key metrics derived from the confusion matrix [41]: Accuracy, Precision, Recall, and F1 Score. Accuracy measures the proportion of correct predictions among all cases. Precision indicates the ratio of correctly predicted positive instances, which is especially important in applications such as medical diagnostics. Recall (or sensitivity) represents the ratio of actual positives that are correctly identified. The F1 Score, the harmonic mean of Precision and Recall, provides a balanced measure of the model's overall performance.

*3.10. Shapley, Linear Discriminant Analysis (LDA) and Principal Component Analysis (PCA)*

In this study, we use SHAP (SHapley Additive exPlanations) to analyze the contribution of statistical and gradient features by assigning each feature an importance value that reflects its effect on the model output, indicating whether it increases or decreases the probability of a given class. Additionally, PCA is applied as an unsupervised method to project data into principal components that capture maximum variance for dimensionality reduction, while LDA is used as a supervised technique to find projections that maximize class separation and minimize within-class variance. Unlike PCA, which focuses on overall variance, LDA enhances class discriminability, making it more effective for classification tasks.

## 4. Results and Discussion

Electrooculography (EOG) data were acquired from twenty healthy volunteers (10 male and 10 female) to ensure balanced representation and minimize demographic bias. All participants had normal or corrected-to-normal vision and no history of neurological or ophthalmological disorders. Experiments were conducted under controlled laboratory conditions at the KUET Biomedical Engineering Laboratory to ensure consistency and reduce environmental interference. Signal acquisition was performed using a BIOPAC MP3X system, selected for its high input impedance and stable amplification characteristics. Disposable Ag/AgCl electrodes with conductive gel were used to maintain low skin–electrode impedance ($< 5$ k$\Omega$), improving signal quality and reducing motion artifacts. Shielded cables further minimized electromagnetic interference.

A two-channel horizontal–vertical configuration with six electrodes was employed: two near the outer canthi for horizontal movements, two above and below one eye for vertical movements, and two reference electrodes on the forehead. This setup enabled effective separation of horizontal and vertical components and reliable detection of both cardinal and diagonal eye movements (Fig. 4). Signals were sampled at 125 Hz, capturing the dominant EOG frequency range (0.1–30 Hz) while maintaining computational efficiency. Participants performed ten eye-movement classes (Stare, Blink, Left, Right, Up, Down, Up-Left, Up-Right, Down-Left, and Down-Right) using a visual reference board with targets positioned approximately 50° from the central axis. Each participant completed 15 trials per class, with each trial lasting approximately 20 seconds, enabling extraction of multiple single-cycle segments. LED-based visual cues and an Arduino-controlled synchronization system ensured consistent timing and reproducibility across trials.

Following acquisition, the raw EOG recordings exhibited baseline drift and high-frequency noise caused by electrode polarization, slow physiological variations, and environmental interference. To enhance signal quality while preserving waveform morphology, a structured two-stage filtering pipeline was implemented.

Initially, a moving average filter with a window size of 30 samples was applied to smooth short-term fluctuations and suppress high-frequency noise. Given the sampling frequency of 125 Hz, this window length provided effective noise reduction while retaining the overall signal trend and essential temporal characteristics required for single-cycle detection.

To mitigate slow baseline drift, a fifth-order Butterworth high-pass filter with a cutoff frequency of 0.2 Hz was subsequently applied using zero-phase filtering to prevent phase distortion and preserve temporal integrity. Following high-pass filtering, any residual DC offset was removed by subtracting the median value of the signal, ensuring improved baseline symmetry and waveform consistency.

The effectiveness of the filtering process was verified through visual inspection of the denoised waveforms, where baseline stabilization and clearer peak structures were observed. These improvements confirmed the suitability of the processed signals for reliable single-cycle detection and subsequent feature extraction.

To support low-latency classification, the processing pipeline was structured to isolate individual EOG movement cycles using polarity-based peak detection followed by fixed-length segmentation. Peak detection was performed to identify distinct signal deflections corresponding to the onset of eye-movement events. Since the polarity of the EOG varies with movement direction, segmentation was conducted separately for positive and negative signal deflections.

Positive peaks were primarily associated with Blink, Up, Right, Up-Right, Up-Left, and Stare movements, while negative peaks corresponded to Down, Left, Down-Right, and Down-Left movements. This polarity-based separation was validated through consistent waveform orientation observed across subjects, where the dominant signal deflection matched the intended gaze direction. This observation enabled reliable cycle identification and improved inter-class separability.

Peak detection was implemented using Python's peak detection algorithm with empirically optimized thresholds: peak height of $\pm 0.10$ V, minimum prominence of $0.13$ V, and minimum peak distance of 140 samples. These parameters were selected to reliably capture true movement cycles while minimizing false detections arising from noise or involuntary micromovements.

Following peak identification, each valid peak was treated as the temporal center of a single EOG cycle. A fixed window of 140 samples on either side of the detected peak was extracted, resulting in a standardized 280-sample segment capturing the complete waveform, including both ascending and descending phases. This symmetric segmentation ensured preservation of the full morphological structure of the movement. For the *Stare* class, which does not produce distinct peaks, fixed-duration segments of equal length (280 samples) were extracted based

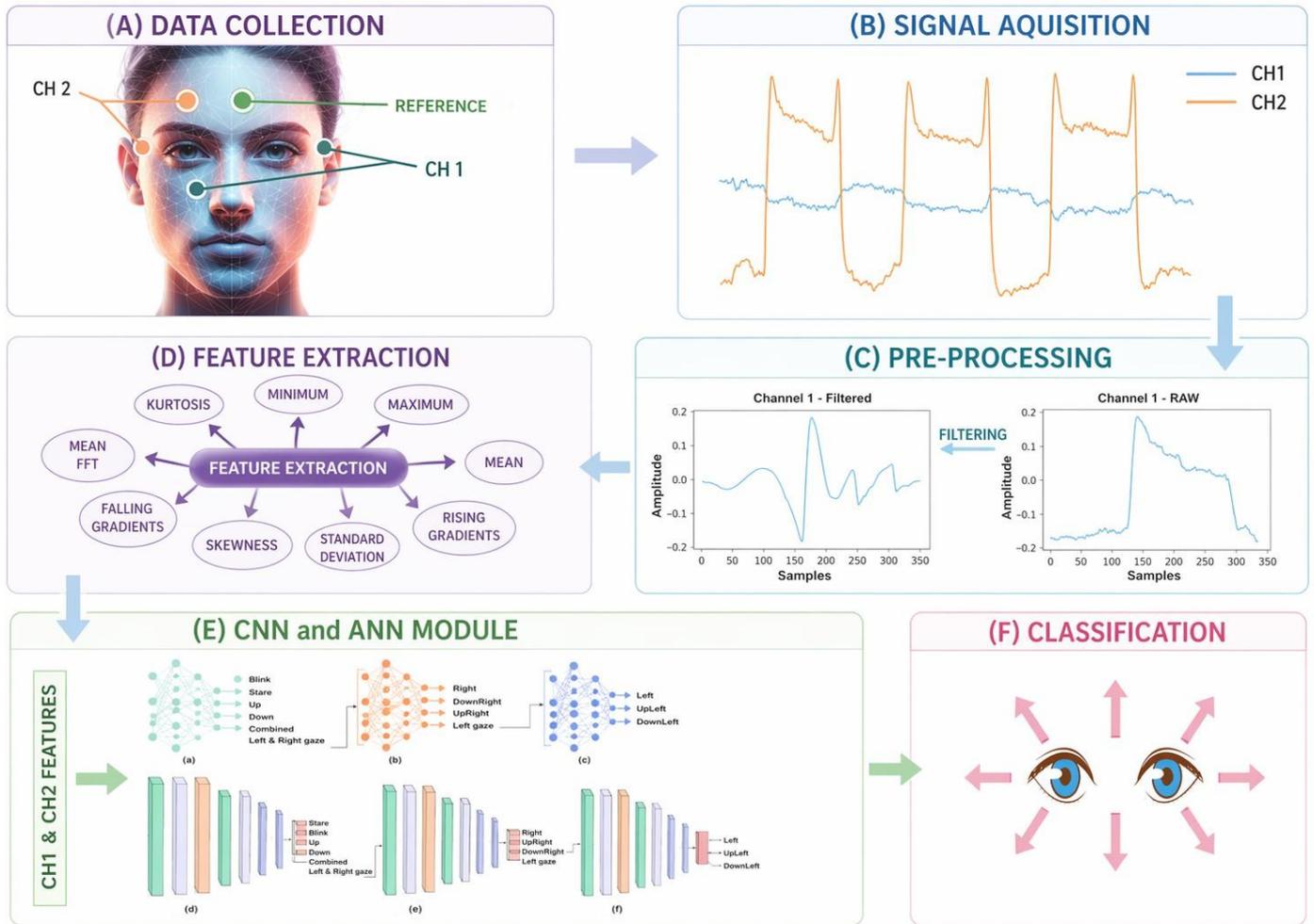

Figure 8: Outline of the experimental procedure.

on temporal alignment to represent steady-state gaze behavior.

From each segmented EOG cycle, 13 features were extracted per channel, resulting in a total of 26 features for the two-channel configuration. The statistical features—maximum, minimum, mean, standard deviation, skewness, kurtosis, and mean Fast Fourier Transform (FFT) magnitude—captured the overall amplitude distribution and frequency characteristics of each channel independently.

To complement these global descriptors, six gradient-based features were computed per channel to quantify the dynamic rate of signal change within each cycle. Gradients were evaluated at normalized amplitude levels (40%, 60%, and 80% of peak magnitude) across the rising and falling phases, enabling characterization of waveform slope variations associated with different eye movements.

The Pearson correlation analysis as shown in Fig.9 of the 26 extracted features revealed structured relationships reflecting both redundancy and complementary dynamics among amplitude- and gradient-based descriptors. Strong correlations were observed among amplitude features within each channel. For Channel 1, maximum and minimum amplitudes showed a strong negative correlation ($r = -0.84$), while maximum amplitude was highly positively correlated with standard deviation ($r = 0.96$) and mean FFT magnitude ($r = 0.89$). Similarly, minimum amplitude exhibited a strong negative correlation with standard deviation ($r = -0.93$). These relationships indicate that higher signal amplitudes are directly associated with increased variability and spectral energy. Gradient-based features demonstrated strong internal consistency, with high correlations among rising (e.g., RG40–RG60: $r = 0.81$) and falling gradients (e.g., FG40–FG60: $r = 0.82$), as well as moderate cross-phase relationships (e.g., RG80–FG60: $r = 0.52$). Comparable patterns were observed in Channel 2, confirming consistent slope dynamics across channels. Cross-channel correlations were generally low (e.g., CH1 Max vs. CH2 Max $\approx 0.09$), indicating that horizontal and vertical channels capture complementary rather than redundant information. In contrast, skewness and kurtosis showed weak correlations with most features ($|r| < 0.35$), suggesting that higher-order statistical descriptors provide additional independent information. Overall, while expected redundancy exists within feature groups, the absence of strong multicollinearity across most feature pairs indicates that the feature set retains sufficient diversity, supporting robust and discriminative classification.

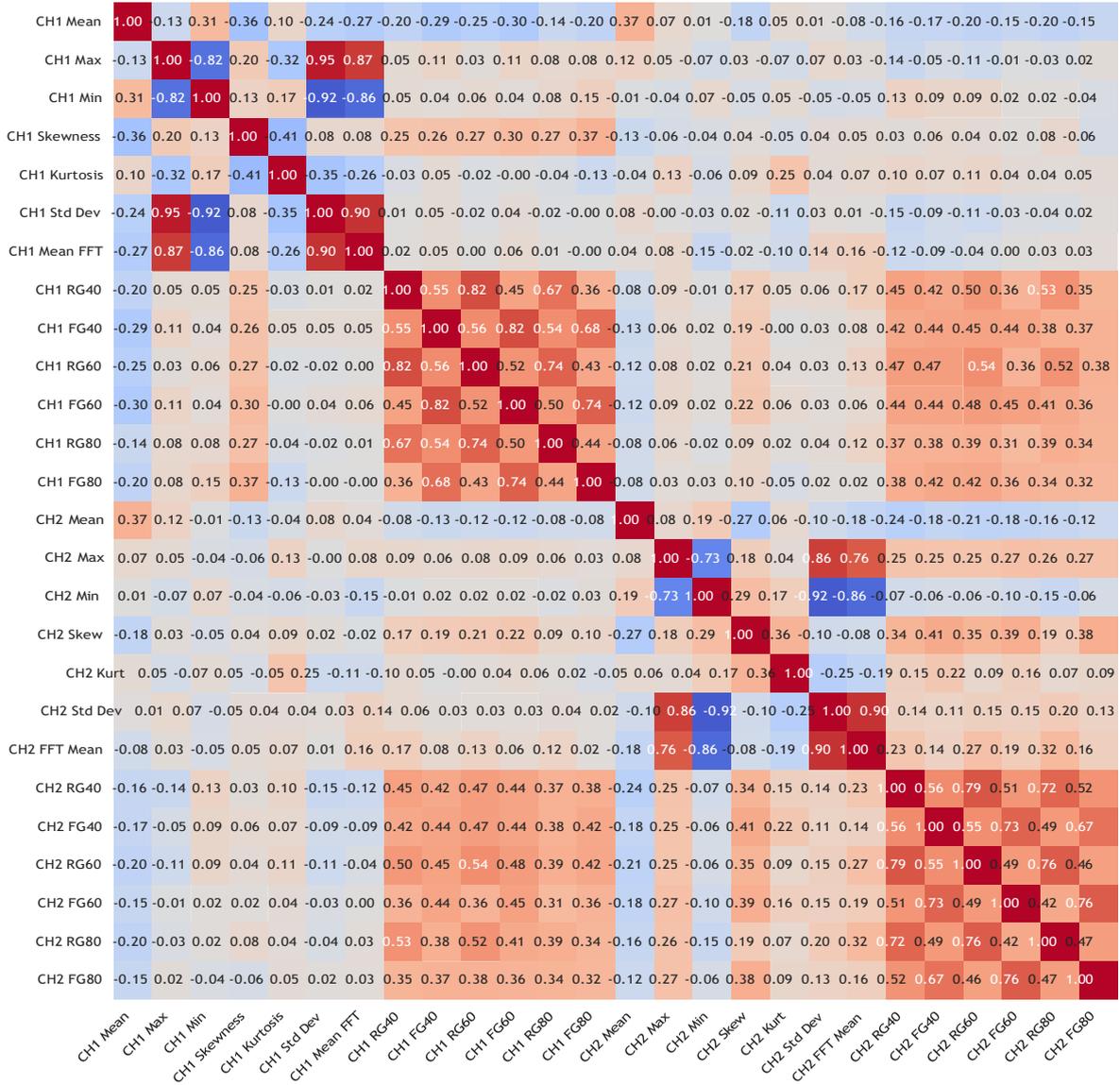

Figure 9: Feature correlation matrix showing inter-feature relationships across the extracted 26-dimensional feature space. Strong positive and negative correlations indicate redundancy and complementary discriminative characteristics among statistical and gradient-based features.

Following feature extraction, class imbalance was observed due to variations in cycle detection across movement categories. To mitigate potential model bias, the Synthetic Minority Oversampling Technique (SMOTE) was applied, resampling each class to 200 instances and producing a balanced dataset across all ten classes. To verify that the synthetic samples preserved the statistical properties of the original data, an independent two-sample $t$-test was performed for each feature within each class. The average $p$-values for all classes were found to be greater than 0.05, indicating that no statistically significant difference was observed ($p > 0.05$) between the original and synthetic feature distributions. This confirms that SMOTE maintained feature integrity while effectively correcting class imbalance. The balanced dataset was then split using stratified sampling into 80% training and 20% testing sets, maintaining equal class proportions. Feature normalization was performed using a Standard Scaler fitted on the training set to prevent data leakage, transforming features to zero mean and unit variance prior to application on the test data. The final dataset consisted of balanced, normalized 26-dimensional feature vectors, which served as input to the ANN and CNN models.

The quantitative evaluation in Table 3 demonstrates high classification accuracy across both standalone and cascaded architectures for single-cycle EOG-based eye-movement recognition. Single-stage (1D) ANN and CNN models achieved strong baseline performance, with the ANN reaching 98.85% accuracy and the CNN 97.35%. The superior ANN performance is attributed to its ability to model global nonlinear relationships within the compact 26-dimensional feature vector, which already encodes statistical, temporal, and spectral characteristics of the EOG signal. Through fully connected layers, the ANN captures inter-feature dependencies without relying on spatial assumptions. In contrast, CNNs operate using convolutional filters with weight sharing, focusing on local feature interactions. This approach is less suited to heterogeneous engineered features, where adjacency does not reflect meaningful relationships, resulting in slightly reduced performance in the standalone setting.

Table 3: Performance comparison of single and cascaded ANN and CNN models for eye-movement classification, including architectural details.

| Model | Stage | Architecture | Classes | Accuracy (%) | Precision (%) | Recall (%) | F1-score (%) |
|---|---|---|---|---|---|---|---|
| ANN | – | FC(26–128–64–32–10) | All classes | 98.85 | 98.89 | 98.85 | 98.85 |
| Cascaded ANN | Stage-1 | FC(26–64–32–10) | Up, Down, Blink, Stare, Combined Left and Right Gaze | 99.50 | 99.51 | 99.50 | 99.50 |
| | Stage-2 | FC(26–64–32–10) | Right, Up-Right, Down-Right, Left Gaze | 99.58 | 99.68 | 99.58 | 99.60 |
| | Stage-3 | FC(26–64–32–10) | Left, Up-Left, Down-Left | 100.00 | 100.00 | 100.00 | 100.00 |
| | Overall | Cascaded (3-stage sequential FC models) | All classes | 99.30 | 99.30 | 99.30 | 99.30 |
| 1D CNN | – | Conv1D(32) → Conv1D(64) → Flatten → FC(64) → FC(10) | All classes | 97.35 | 97.48 | 97.35 | 97.33 |
| Cascaded CNN | Stage-1 | Conv1D(32) → Conv1D(64) → Flatten → FC(64) → FC(10) | Up, Down, Blink, Stare, Combined Left & Right | 99.55 | 99.56 | 99.55 | 99.55 |
| | Stage-2 | Conv1D(32) → Conv1D(64) → Flatten → FC(64) → FC(10) | Right, Up-Right, Down-Right, Left Gaze | 99.42 | 99.42 | 99.42 | 99.42 |
| | Stage-3 | Conv1D(32) → Conv1D(64) → Flatten → FC(64) → FC(10) | Left, Up-Left, Down-Left | 100.00 | 100.00 | 100.00 | 100.00 |
| | Overall | Cascaded (3-stage CNN pipeline) | All classes | 99.75 | 99.76 | 99.75 | 99.76 |

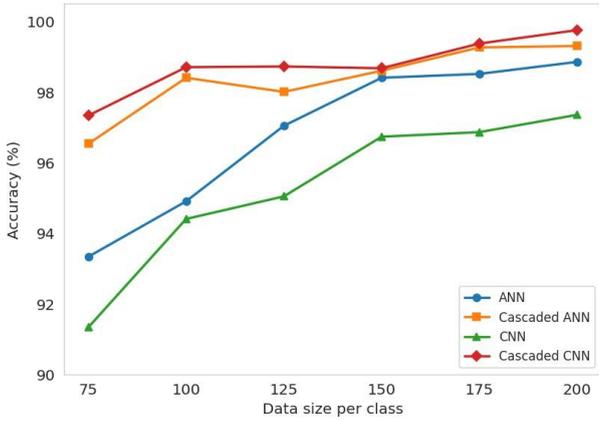

Figure 10: Accuracy comparison of ANN, cascaded ANN, CNN, and cascaded CNN with varying data size per class.

To improve performance, cascaded architectures were introduced, decomposing the 10-class problem into sequential subtasks. This hierarchical approach reduced inter-class ambiguity and enhanced classification accuracy, with the cascaded ANN achieving 99.30% and the cascaded CNN reaching 99.75%. Stage-wise results show consistently strong performance, with Stage-1 achieving 99.50% (ANN) and 99.55% (CNN), Stage-2 maintaining similar accuracy, and Stage-3 achieving 100% across all metrics.

The improvement arises from progressive reduction of classification complexity, where each stage operates on a smaller, more homogeneous subset, enabling clearer decision boundaries and improved separability. This staged framework also enhances CNN performance; while CNNs are less effective in the standalone setting, they benefit from reduced class complexity in later stages, where convolutional filters can better capture subtle feature interactions. The cascaded mechanism further refines classification through sequential narrowing of decision space, culminating in near-perfect separability in Stage-3, where classification is limited to left-oriented movements and depends primarily on vertical variation.

Principal Component Analysis (PCA) plot in Fig. 11 was performed to examine class separability in the reduced feature space. The first two components explain 27.40% (PC1) and 15.76% (PC2) of the total variance, yielding a cumulative 43.16%. Despite capturing less than half of the variance, clear structural organization among ocular movement classes is evident.

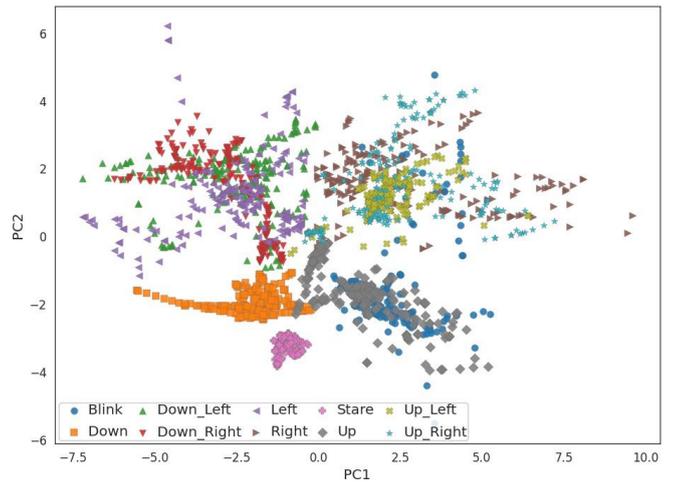

Figure 11: Principal Component Analysis (PCA) visualization of the extracted 26-dimensional feature space projected onto the first two principal components. The distribution illustrates class separability and underlying feature structure across different eye-movement categories.

PC1 primarily reflects horizontal movement dynamics, with rightward-related classes (Right: 2.34, Up-Right: 2.29, Blink:

2.43) in positive regions and left-oriented movements (Left: $-2.68$, Down-Left: $-2.63$, Down-Right: $-2.82$) in negative regions. PC2 captures vertical and waveform characteristics, where the Stare class shows strong separation ($-3.38$), distinguishing steady-state gaze from dynamic movements. Vertical movements (Up, Down) lie in negative regions, while diagonal and lateral classes extend into positive PC2 space, reflecting mixed directional influences.

Cardinal movements form compact clusters, whereas diagonal classes exhibit partial overlap (e.g., Down-Left and Down-Right), indicating limitations of linear separability. Overall, PCA confirms that the extracted features encode meaningful directional structure while revealing inter-class proximity, supporting the use of nonlinear and hierarchical models for improved classification.

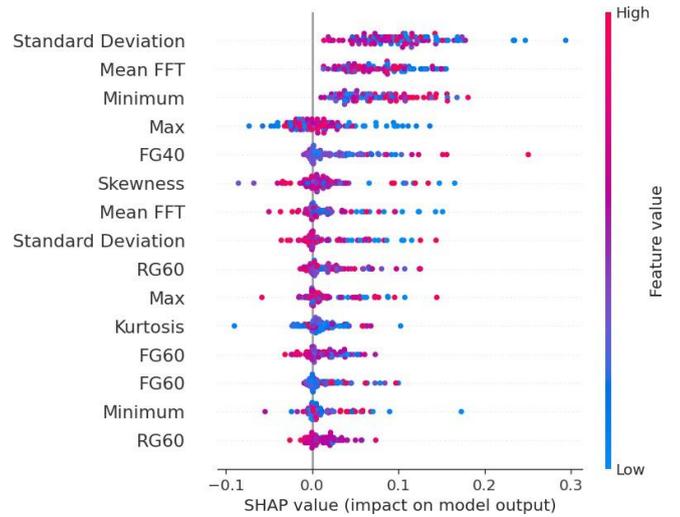

Figure 13: SHAP summary plot showing the relative contribution of extracted EOG features to model prediction. Feature color represents the feature value, while the horizontal axis indicates the SHAP value and its impact on the model output.

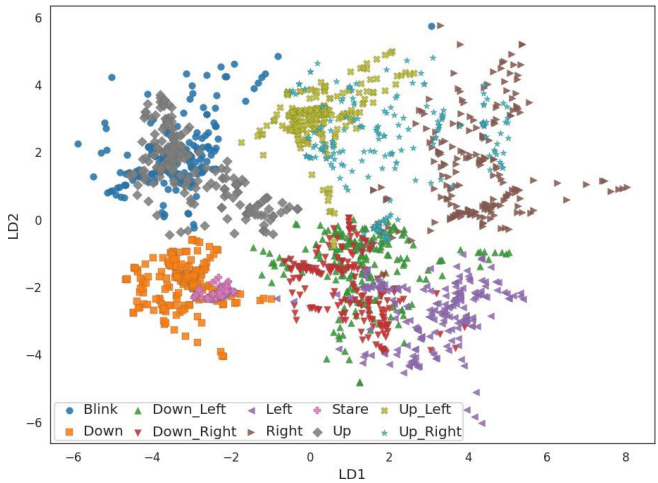

Figure 12: Linear Discriminant Analysis (LDA) projection of the 26-dimensional feature space onto the first two discriminant components. The visualization illustrates enhanced class separability compared to PCA due to supervised variance maximization between eye-movement categories.

Linear Discriminant Analysis (LDA) as shown in Fig. 12 was performed to evaluate supervised class separability by maximizing between-class variance and minimizing intra-class dispersion. Compared to PCA, the LDA projection shows markedly improved separation among the ten ocular movement classes. Cardinal directions (Up, Down, Left, Right) form compact and well-defined clusters, indicating strong discriminative encoding within the 26-dimensional feature space. The Stare class is tightly grouped and clearly isolated from dynamic movements, reflecting its steady-state characteristics. Right-oriented movements occupy the positive LD1 region, while left-oriented movements lie oppositely, demonstrating effective horizontal discrimination. Although diagonal classes (e.g., Up-Left, Down-Right) remain relatively closer due to shared directional components, inter-class boundaries are significantly clearer than in PCA space. This improved separability aligns with the high classification accuracy of the ANN and cascaded models, confirming that the extracted features provide a robust discriminative structure for hierarchical multi-class classification.

The SHAP summary plot in Fig. 13 highlights the relative importance and influence of the extracted features on model predictions. Features such as standard deviation, mean FFT, and minimum amplitude exhibit the highest impact, indicating that signal variability, spectral content, and extrema are the most discriminative for eye-movement classification. Higher values of these features generally contribute positively to the model output, while maximum amplitude and skewness show mixed contributions, suggesting class-dependent behavior. Gradient-based features, including FG40, FG60, and RG60, together with higher-order statistical features such as kurtosis, contribute moderately by capturing finer signal characteristics. Overall, the SHAP analysis confirms that the combination of statistical, spectral, amplitude-based, and gradient-based features effectively represents the underlying EOG dynamics, supporting the high classification performance achieved in the proposed framework.

Latency analysis in Table 4 further highlights the trade-offs between architectural design and computational efficiency. Latency values represent end-to-end processing time per sample, including both preprocessing and classification. To better reflect real-time deployment, inference latency was measured using direct model invocation rather than higher-level prediction APIs, thereby reducing framework overhead. The standalone ANN demonstrated the lowest inference latency (38.6 ms), followed by the standalone CNN (50.82 ms), reflecting the lightweight nature of single-stage models. In contrast, cascaded architectures incur higher end-to-end latency due to sequential model execution. However, a key distinction emerges from their internal computation mechanisms. ANN-based models rely on dense fully connected operations, where each neuron processes all input features, resulting in a higher number of parameters and increased computational cost, particularly when executed across multiple cascade stages. Conversely, CNNs employ parameter sharing through convolutional filters, significantly reducing the number of trainable parameters and enabling more efficient computation. This advantage is reflected in the lower overall latency of the cascaded CNN (82.25 ms) compared to the cascaded ANN (99.72 ms). Overall, the results reveal a complementary relationship between architecture and problem formulation. ANN is effective for modeling global feature interactions in single-stage classification, whereas CNN benefits from parameter sharing and localized feature extraction, performing better within the cascaded framework. Notably, the cascaded CNN achieves the best balance between accuracy and latency, making it the most suitable approach. These findings confirm that accurate multi-class classification using single-cycle EOG signals is achievable through structured feature extraction and hierarchical, stage-wise learning.

Table 4: Latency comparison across eye-movement classes.

| Class | ANN | Casc. ANN | 1D CNN | Casc. CNN |
|---|---|---|---|---|
| Blink | 27.96 | 47.15 | 37.85 | 42.08 |
| Down | 23.27 | 48.77 | 90.38 | 59.01 |
| Down-Left | 34.87 | 95.32 | 30.93 | 161.74 |
| Down-Right | 35.79 | 71.99 | 34.17 | 92.00 |
| Left | 23.50 | 137.66 | 83.02 | 123.37 |
| Right | 25.50 | 87.56 | 58.10 | 79.07 |
| Stare | 37.70 | 29.22 | 23.72 | 45.53 |
| Up | 24.11 | 47.88 | 45.83 | 53.70 |
| Up-Left | 28.51 | 109.79 | 35.01 | 120.50 |
| Up-Right | 37.63 | 80.52 | 33.02 | 88.20 |
| Stage-1 | – | 36.03 | – | 42.49 |
| Stage-2 | – | 37.04 | – | 41.33 |
| Stage-3 | – | 35.31 | – | 44.18 |
| Overall | 38.6 | 99.72 | 50.82 | 82.25 |

## 5. Conclusion

The objective of this study was to develop a robust strategy to capture a wide range of eye movements faster than human reaction time from single-cycle EOG information. To achieve this, a complete processing pipeline was established, beginning with data acquisition and signal conditioning, followed by feature extraction and model development. The feature engineering process demonstrated that gradient-based descriptors play a critical role in capturing the temporal dynamics and morphological characteristics of single-cycle EOG signals, enabling improved discrimination among classes with subtle variations, particularly the diagonal movements. To circumvent the complexity of single-cycle multi-class classification, we explored both 1D and cascaded architectures of ANN and CNN. The later algorithm exhibited the best overall performance achieving an accuracy of 99.75%, precision of 99.76%, recall of 99.75%, F1-score of 99.76%, and system of 82.25 ms, which is significantly lower than the average human reaction time, thereby ensuring suitability for real-time applications. The introduction of the latency-aware figure-of-merit further validated the effectiveness of the proposed approach in balancing classification performance with real-time responsiveness. The results confirm that hierarchical (cascaded) learning strategies are highly effective in handling complex multi-class EOG classification tasks, especially under the constraints of limited data availability. Overall, the proposed system demonstrates that high-accuracy, low-latency EOG-based interaction is feasible using minimal signal duration, making it highly suitable for practical deployment in real-time human-computer interactions, smart wearable devices, and assistive technologies.